\documentclass[npg]{copernicus2}

\usepackage{ulem}

\begin{document}


\title{Turbulence in the Interstellar Medium}

\author[1,2]{D. Falceta-Gon\c calves}
\author[2]{G. Kowal}
\author[3]{E. Falgarone}
\author[4,5,6]{A. C.-L. Chian}

\affil[1]{SUPA, School of Physics \& Astronomy, University of St Andrews, North Haugh, St Andrews, Fife KY16 9SS, UK}
\affil[2]{Escola de Artes, Ciencias e Humanidades, Universidade de Sao Paulo, Rua Arlindo Bettio 1000, CEP 03828-000, Sao Paulo, Brazil}
\affil[3]{LERMA/LRA, CNRS, Ecole Normale Sup\'erieure and Observatoire de Paris, 24 rue Lhomond, 75231 Paris Cedex, France}
\affil[4]{Observatoire de Paris, LESIA, CNRS, 92190 Meudon, France}
\affil[5]{National Institute for Space Research (INPE) and World Institute
for Space Environment Research (WISER), P. O. Box 515,
12227-010 S\~ao Jos\'e dos Campos-SP, Brazil}
\affil[6]{School of Mathematical Sciences, University of Adelaide, Adelaide, SA 5005, Australia}


\runningtitle{Interstellar turbulence}

\runningauthor{Falceta-Gon\c calves et al.}

\correspondence{Diego Falceta-Gon\c calves\\ (dfalceta@usp.br)}

\received{}
\pubdiscuss{} 
\revised{}
\accepted{}
\published{}


\firstpage{1}

\abstract{
Turbulence is ubiquitous in the insterstellar medium and plays a major
role in several processes such as the formation of dense structures and
stars, the stability of molecular clouds, the amplification of magnetic
fields, and the re-acceleration and diffusion of cosmic rays.  Despite its importance,
interstellar turbulence, alike turbulence in general, is far from being fully understood. 
In this review we present the basics of turbulence physics,
focusing on the statistics of its structure and energy cascade.  We explore
the physics of compressible and incompressible turbulent flows, as well as 
magnetized cases.  The most relevant observational techniques that provide 
quantitative insights of interstellar turbulence are also presented. We 
also discuss the main difficulties in developing a three-dimensional view 
of interstellar turbulence from these
observations.  Finally, we briefly present what could be the the main sources 
of turbulence in the interstellar medium.
\keywords{turbulence, interstellar medium, plasmas, magnetic fields, shocks}}

\maketitle  



\section{Introduction: the basics of turbulence}

Turbulence is characterized by chaotic motions in a fluid 
\citep{rempel04, he05, chian06, chian07, chian10} that lead to
diffusion of matter and dissipation of kinetic energy.  It is to be 
stressed that not all chaotic motions in a fluid may be called ``turbulent".  
Because of its chaotic nature
turbulence can only be studied and modelled in terms of statistical quantities.
Long-term deterministic local properties of a turbulent fluid are unpredictable.

For nearly incompressible and unmagnetized fluids, the temporal evolution of the
fluid velocity field is given by the Navier-Stokes equation:

\begin{equation}
\frac{\partial {\bf u}({\bf x},t)}{\partial t} + {\bf u}({\bf x},t) \cdot {\bf \nabla u}({\bf x},t) = - \frac{{\bf \nabla}p({\bf x},t)}{\rho({\bf x},t)}+\nu \nabla^2 {\bf u}({\bf x},t)+{\bf F}({\bf x},t),
\label{eq1}
\end{equation}

\noindent
where ${\bf u}({\bf x},t)$ represents the velocity field, $p$ the pressure,
$\nu$ the kinematic viscosity, and ${\bf F}$ an external force
normalized by the local density.  $\rho$ is the gas mass density and is set 
constant in the incompressible case (with ${\bf \nabla \cdot u}=0$).  Even in
this simplified mathematical description the fluid dynamics is not a trivial
solution.  Equation~\ref{eq1} is non-linear, as seen from the advective term in
the left hand side, and non-local - in the sense that the local properties of the
fluid are related to all the other regions -, through the pressure term.  The
incompressibility condition results in an infinite sound speed, and in an
instantaneous propagation of any perturbation in the fluid.

\citet{burg39} modeled the time evolution of the simplified version of the
Navier-Stokes equation by considering $\nabla p = 0$.  This equation has exact
solutions, which may sound interesting, but it results in non-universal
``turbulence". Eventhough Burgers turbulence models have gained increasing
interest due to their ability to describe the statistics of shock induced
structures, and many other applications \cite[see review by][]{bec07}.

In the full Navier-Stokes equation, perturbations in ${\bf u}({\bf x},t)$ are
expected to have their distribution changed due to non-linear terms.  These
instabilities may drive local vorticity and result in the fragmentation of large
amplitude eddies into smaller ones, creating a turbulent pattern.  As imagined
by \citet{ric22}, {\it big whirls have little whirls that feed on their
velocity, and little whirls have lesser whirls, and so on to viscosity}.  This
statement represents one of the first conceptual descriptions of the energy
cascade in turbulent flows.  The shear drives unstable motions at large scales,
which are broken and fragmented into smaller
vortices, down to the smallest scales where they are damped, e.g. due to
viscosity.  In an incompressible viscous fluid this damping scale is
that at which the timescale for
viscous damping is of the order of the turnover dynamical time.
At that scale, the eddy kinetic energy is transferred to internal energy due to viscosity.
Turbulence is naturally developed over larger range of scales if
viscosity is small, i.e. with large Reynolds number ($Re = UL/\nu \gg 1$), being the
characteristic velocity $U$ injected at a lengthscale $L$.

\citet[hereafter K41]{kol41} realized that it would be possible to
solve the Navier-Stokes equation for a turbulent flow if ${\bf u}({\bf
  x},t)$ is considered a stochastic distribution.  One of the key
assumptions in the K41 theory is that the energy transfer rate $\epsilon$ 
should be constant at all scale. It is defined as $\epsilon \simeq \delta
u_l^2/\tau_l$, where $\delta u_l$ is the velocity fluctuation
amplitude at lengthscale $l$, and $\tau_l = \tau_{\rm eddy} =
l/\delta u_l$ its dynamical timescale\footnote{Note that we 
  distinguish $\tau_l$ and $\tau_{\rm eddy}$ here, since $\tau_l$
  represents the timescale for energy transfer at scale $l$, while
  $\tau_{\rm eddy}$ is the eddy turnover timescale. In the K41 theory
  both timescales are the same, but this is not true for other cases,
  e.g. as in some magnetized cases}.  Therefore, one obtains:

\begin{equation}
\delta u_l \simeq (\epsilon l)^\frac{1}{3}.
\label{eq2}
\end{equation}

Equation~\ref{eq2} means that turbulence can be modeled
by scaling laws.  This would be true within the so called 
{\it inertial range of scales}, i.e. the scales where the energy transfer
rate is constant, generally between the energy injection and the
viscous damping scales.  The velocity power spectrum $P_u(k)$ is
defined\footnote{The power spectrum is defined as the one
  dimensional spectrum in Fourier space while the energy spectrum,
  generally defined as $E_u(k) =k^2P_u(k)$ is the three-dimensional
  spectrum. For the sake of simplicity, we use the term {\it power
    spectrum} to represent the latter.} here by $\int_{k=1/l}^\infty
P_u(k')dk'= \delta u_l^2$, from which we obtain the standard Kolmogorov
power spectrum for the velocity field:

\begin{equation}
P_u(k) \propto \epsilon^{2/3} k^{-5/3}.
\end{equation}

In other words, it is possible to reinterpret Kolmogorov's idea in Fourier space
in terms of non-linear interaction between similar wavenumbers.  This theory is a
result of the so-called {\it locality}, i.e. similar wavenumbers,
$k=2\pi/\lambda$, of the non-linear wave-wave interaction that result in the
energy cascade through smaller scales \cite{kra65a}.  From the spectral form of
the Navier-Stokes equation, the three-wave interactions follow the selection
rule $k_3=k_1+k_2$.  The extrema are found at $k_3\rightarrow0$ and $k_1=k_2$,
which is the locality assumed in Kolmogorov's theory, resulting in $k_3=2k_1$.
 
The theory also predicts the scaling laws for the moments of velocity spatial
increments, known as {\it velocity structure functions}, defined as:

\begin{equation}
S_p(l)=\left\langle \left\{\left[ \textbf{u}\left( \textbf{r} + \textbf{l} \right) - \textbf{u} \left( \textbf{r} \right) \right] \cdot \textbf{l}/l \right\} ^p \right\rangle = C(p) \epsilon^{p/3}l^{p/3},
\end{equation}

\noindent
where $p$ is a positive integer representing the moment order and ${\bf l}$ is
the spatial increment vector.  In incompressible fluids, if the turbulence is
considered {\bf \it homogeneous}, {\bf \it isotropic} and {\bf \it
self-similar}, i.e. scale invariant, then:

\begin{equation}
S_p(l)=C(p) \epsilon^{p/3} l^{p/3} ,
\end{equation}

\noindent
where $C(p)$ was initially assumed by Kolmogorov to be constant with $p$.

One of the main successes of the Kolmogorov-Obukhov turbulence theory is
the explanation of  
the empirical determination of the diffusion coefficient by
\citet{ric26}, done more than a decade before K41.  The diffusion coefficient is
related to the time evolution of the separation between Lagrangian points (e.g.
particles dragged by the flow) in a turbulent medium.  The probability
distribution function $\Phi$ of pairs of points separated by a
distance ${\bf r}$ may be described as:

\begin{equation}
\frac{\partial \Phi \left({\bf r},t\right)}{\partial t} = \frac{1}{r^2} \frac{\partial}{\partial r} r^2 K(r) \frac{\partial \Phi \left({\bf r},t\right)}{\partial r},
\label{eq6}
\end{equation}

\noindent
where $K(r)$ represents the diffusion coefficient.  It is easy to determine,
from dimensional analysis, that if $\dot{r} = u(r) \propto r^{1/3}$ as in the
Kolmogorov scaling, the diffusion coefficient for the inertial range will be
$K(r) = k_0 \epsilon^{1/3} r^{4/3}$, the scaling proposed by \citet{ric26}.
This diffusion coefficient for the inertial range substituted in
Equation~\ref{eq6} then results in:

\begin{equation}
\Phi \left({\bf r},t\right) = \frac{A}{(k_0 t)^3\epsilon} \exp\left(-\frac{9r^{2/3}}{4k_0\epsilon^{1/3}t} \right),
\end{equation}

\noindent
where $A$ is a normalization coefficient.  The Richardson
distribution is therefore non-Gaussian. Several experiments and numerical
models have shown the validity of the turbulent diffusion scaling \cite{ell96,
fun98, zov94, bof02}, as has also been recently used in the predictions of
stochastic magnetic reconnection \footnote{this term accounts for the magnetic reconnection that is induced by turbulent motions near the current sheet - separation layer between fields with components of opposed directions -, which would then result in reconnection rates as a function of the stochastic motions of the fluid.} \cite{laz12}.

\begin{figure}[t]
\vspace*{2mm}
\begin{center}
\includegraphics[width=8.3cm]{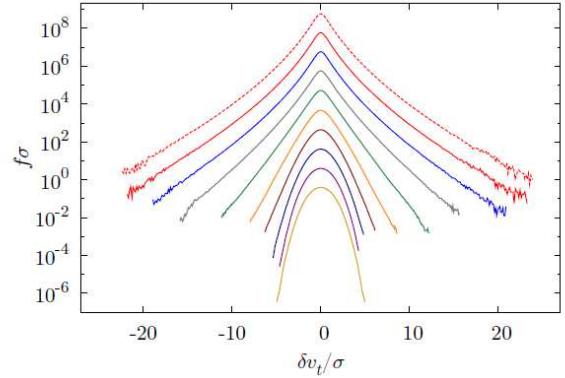}
\end{center}
\caption{PDF of velocity increments as a function of the lag length $\left|{\bf
l}\right|$, from small (top) to large scales (bottom) \cite[extracted
from][]{wil10}.  The non-Gaussianity is clear for velocity increments at small
scales. \label{fig1}}
\end{figure}

This theory of turbulence has been quite successful in reproducing most of
experimental data, and there is a flourishing literature with hundreds of works available
e.g. \citet{arm95, lea98, bal05, koga07, bou09, chian09, che10, sah10, chian11, 
cha12, hur12, miranda13}, just to mention a few.  Naturally, many authors criticized the fact of
$C(p)$ is a constant in Kolmogorov's initial theory, given the breakdown of self-similarity
at small scales and the possible non-universality of turbulence (given its
``memory" related to the energy injection).  These criticisms have been later addressed in
the Kolmogorov-Obukhov  turbulence theory \cite{kol62, obu62}, including the
effects of {\it intermittency}.  Intermittency results from rare and large local fluctuations 
in the  velocity field which break the  similarity condition \cite{fri95}.  
One of the effects of intermittency is observed in the
probability distribution function (PDF) of velocity longitudinal increments $\delta u_l = [{\bf
u}({\bf r}+{\bf l})-{\bf u}({\bf r})]\cdot \hat{l}$, which shows large deviations
from the Gaussian distribution at small scales, with large amplitude tails and
peaked distributions at $\delta u_l \sim 0$ (see Figure~\ref{fig1}).
\citet{kra91} pointed that sharp shocks could, for intance, result in 
more regions with smooth fluid flows and also more regions with sharp transitions 
in velocities, compared to the standard picture of the self-similar K41 turbulence. 
We would then expect non-Gaussian PDFs at both small and large scales.

Many authors attempted to theoretically determine the scalings of turbulence
with intermittency.  One of the most successful approaches is the multifractal
description for the energy dissipation field proposed by \citet{she94}.  This
theory results in $S_p(l) \propto l^{\zeta(p)}$, with:

\begin{equation}
\zeta(p)=\frac{p}{3}(1+\frac{2}{3})+(3-D')\left[1-\left(1-\frac{2}{3(3-D')} \right)^{p/3} \right],
\end{equation}

\noindent
where $D'$ represents the dimensionality of the dissipation structures.
In the 
Kolmogorov-Obukhov theory, structures of highest dissipation are filamentary, better described
then by $D' \sim 1$, while recent numerical simulations reveal a dominance
of two dimensional intermittent structures at small scales
\cite[e.g.][]{moi04,kow07a,kow07b,bol12}, what is also supported by 
experimental data \cite[e.g.][]{fre03, the07}. Multifractal analysis 
of Voyager 1 and 2 {\it in situ} data 
 have also showed intermittent features 
on the magnetic turbulence at the solar wind and the termination shock \citep{macek08,macek11,macek12}.
On the theoretical side, \citet{bir13} derived a 
statistical solution of the stochastic Navier-Stokes equation from the
linear Kolmogorov-Hopf differential equation, accounting for the She-Lev\^eque
intermittency corrections. His results satisfactorily reproduce the PDFs built
on observations and numerical simulations of turbulent flows.
Compressibility and coupling between magnetic fields and the plasma
flow - both present in the dynamics of the interstellar medium
(ISM) - make the description of the interstellar turbulence even more
complex.

\subsection{Supersonic turbulence}
\label{sec1.1}

Compressible plasmas are of great interest in astrophysics, and particularly in the case
of interstellar turbulence.
Compressibility in turbulent flows results in the formation of a
hierarchy of density structures, viewed as dense cores nested in less
dense regions, which are in turn embedded in low density regions and so on. Such a
hierarchical structure was described by \citet{von51} as:

\begin{equation}
\frac{\rho_\nu}{\rho_{\nu-1}} = \left(\frac{l_\nu}{l_{\nu-1}} \right)^{-3\alpha},
\end{equation}

\noindent
where $\rho_\nu$ represents the average density of a structure at hierarchical
level $\nu$, at a lengthscale $l$, and $\alpha$ the compressibility degree, assumed to be the 
same at each level.  The dimensionality of the system is obtained by $D'=3-3\alpha$.
Therefore, the average mass within each substructure must follow the relation
$M_l \propto l^{3-3\alpha}$.

The density hierarchy as described above must then be coupled to the local
turbulent motions.  The energy density transfer rate must now be rewritten as
$\epsilon_l = \rho_l \delta u_l^3/l$ to account for the density changes at
different scales \cite{lig55}.  If, once again, one assumes the 
constancy of the energy transfer rate across scales within the inertial range
\cite{fle96}, one obtains the scaling of the amplitude of the velocity fluctuations:

\begin{equation}
\delta u_l \propto l^{\frac{1}{3}+\alpha},
\end{equation}

\noindent
and the velocity power spectrum is then given by:

\begin{equation}
P_u(k) \propto k^{-5/3 -2\alpha}.
\end{equation}

Note that for stationary energy distribution solutions in compressible
turbulence, $\alpha> 0$ which results in steeper velocity power spectra, compared to
the standard K41 scaling.  The density power spectrum, on the other hand,
instead of following the velocity field as a passive scalar would do, presents a distinct
power spectrum given by:

\begin{equation}
P_\rho(k) \propto k^{6\alpha-1},
\end{equation}

\noindent
i.e. for $\alpha \sim 1/6$, the power spectrum of the density field becomes flat
in the inertial range.
One of the most striking results of the hierarchical model for the density field
in compressible turbulence is its ability to recover the standard Kolmogorov
scalings for the density weighted velocity field ${\bf v} \equiv \rho^{1/3} {\bf
u}$ \cite{fle96}.

Numerical simulations of compressible turbulence have confirmed the scalings
described above for $\alpha \simeq 0.15$ \cite{kri07, kow07b}, close to $\alpha
= 1/6$ for which the density power spectrum becomes flat.  The velocity power
spectrum on the other hand becomes $P_u(k) \propto k^{-2}$.  Remarkably, this is
the exact slope obtained for Burger's turbulence, despite the different
framework of that theory.

\subsection{Magnetized turbulence}

Magnetic fields introduce further complexity in the plasma 
dynamics that can be described by the magneto-hydrodynamic (MHD)
equations in the fluid approximation and assuming perfect coupling between the field and 
the plasma:

\begin{eqnarray}
\frac{\partial {\bf u}}{\partial t} + {\bf u} \cdot {\bf \nabla u} = - \frac{{\bf \nabla}p}{\rho}+ \nu \nabla^2 {\bf u}+\frac{\left( {\bf \nabla \times B} \right) \times {\bf B}}{4\pi\rho}+{\bf F},
\label{eq13}
\end{eqnarray}

\begin{equation}
\frac{\partial {\bf B}}{\partial t} = {\bf \nabla \times} \left( {\bf u}{\bf \times B} \right)+\eta \nabla^2{\bf B},
\label{eq14}
\end{equation}

\noindent
where ${\bf B}$ is the magnetic field and $\eta$ the plasma resistivity
($\eta =0$ for ideal plasmas).

Let us first consider an external uniform magnetic field $B_0$.
Any perturbation in the fluid velocity field will be coupled to the magnetic
field. The magnetic tension/pressure results in a decrease of the non-linear
growth of perturbations, but only of those perpendicular to the magnetic field
lines. This complex coupling between the flow and magnetic field makes the
modelling of turbulence in magnetized plasmas an interesting task\footnote{More
details on MHD turbulence may be found in \citet{bis03}}.

\subsubsection{The Iroshnikov-Kraichnan model}

An useful simplification to the equations above is made by 
considering ${\bf B}={\bf B}_0+{\bf \delta B}$, and using the 
Els\"{a}sser variable ${\bf z^\pm}={\bf
u}\pm{\bf \delta \breve{B}}$, where $\breve{B} = B/(4\pi \rho)^{1/2}$.  This has
been independently derived by \citet{iro63} and \citet[]{kra65a, kra65b} (IK
hereafter).  From this change of variables, Eqs~\ref{eq13} and \ref{eq14} result
in \cite[see][]{sch07}:

\begin{equation}
\frac{\partial {\bf z^\pm}}{\partial t} \mp v_A \nabla_{||}{\bf z^\pm}+{\bf z^\mp} \cdot {\bf \nabla z^\pm}  = -{\bf \nabla} p + \frac{\nu + \eta}{2} \delta z^\pm + \frac{\nu - \eta}{2} \delta z^\mp +{\bf F},
\end{equation}

\noindent
where $v_A = B_0/\sqrt{4 \pi \rho}$ is the Alfv\'en velocity 
and $\nabla_{||}$ is the spatial derivative parallel to
the direction of the mean magnetic field.

In their model, Iroshnikov and Kraichnan proposed that incompressible magnetized turbulence
results from the non-linear interactions of counter propagating waves packets.
The timescale for the two wave packets 
to cross each other is of order of the Alfv\'en
time $\tau_A \sim l_{||}/v_A$, where $l_{||}$ is the lengthscale of the wave packet 
parallel to the mean magnetic field.  In their phenomenological description of the MHD
turbulence, the interactions between the wave packets 
are {\bf weak}, i.e. $|{\bf z}^\pm|
\ll \breve{B}_0$ or the field perturbations are much smaller than $B_0$.
Notice that, superimposed to the magnetic fluctuations, the fluid is also perturbed 
and the dynamical timescale of a fluid ``eddy'' is $\tau_{\rm eddy} 
\equiv l/ \delta u_l$. The different wave modes (mechanical and magnetic 
perturbations) thus interact with each other.
For the interaction between modes to be weak the Alfv\'en time must be much
smaller than the dynamical timescale, i.e.
$\tau_A \ll \tau_{\rm eddy}$.  The non-linear decay of the wave packets
in such weak interactions, and subsequently the turbulent cascade, can only
occur after several interactions.  Since interactions are random, the wave packet
amplitude changes in a random walk fashion, i.e. $N = (\tau_l/\tau_A)^{1/2}$
interactions are needed for the wave packet to significantly change. At the same time, 
$N$ is also defined by the number of crossings in a decay timescale 
$N = \tau_l / \tau_{\rm eddy}$, which results in:

\begin{equation}
\tau_l \sim \frac{\tau_{\rm eddy}^2}{\tau_A} \sim \frac{l^2 v_A}{l_{||} \delta u_l^2}.
\end{equation}

Therefore, the turnover time at scale $l$ is longer by a large factor
and, as expected, the non-linear cascade proceeds much more slowly.

The second major assumption in the IK theory of weak turbulence
is its isotropy, i.e. $l_{||} \sim
l$. Substituting this scaling into the relation $\epsilon = \delta u_l^2/\tau_l$, one obtains:

\begin{equation}
\delta u_l \sim (\epsilon v_A)^{1/4} l^{1/4}\\ {\rm and}\\
P_u(k) \sim (\epsilon v_A)^{1/2} k^{-3/2}.
\end{equation}

There is evidence for an IK cascade in the solar wind and
interplanetary medium \cite[e.g.][]{bam08, ng10}.  However, many observations of
the solar wind turbulence also suggest a more Kolmogorov-like turbulence, i.e.
$\propto k^{-5/3}$ \cite[e.g. the early studies of \citeauthor{col68},
\citeyear{col68}; \citeauthor{mat82}, \citeyear{mat82}; and the more recent
papers by][]{ale08, chian09, sah10, li11, chian11, koz12, hel13}.  It is possible though
that a mix of both cascades may occur, as pointed by e.g. \citet{sal09} and
\citet{ale13}, which showed a mix of K41 and IK cascades for the magnetic and
velocity field fluctuations, respectively.  Moreover, most of these data also
reveal the solar wind turbulence to be highly anisotropic (i.e. $\delta u_l^{||}
\neq \delta u_l^{\perp}$) with respect to the local magnetic field \citep{hor08,hor12}.

As pointed by \citet{gol01}, one of the main issues raised by the solar
wind is {\it why is the power spectrum of this anisotropic, compressible,
magnetofluid often Kolmogorov-like?}

\subsubsection{The Goldreich-Sridhar model}

\begin{figure*}[ht]
\vspace*{2mm}
\begin{center}
\includegraphics[width=5cm]{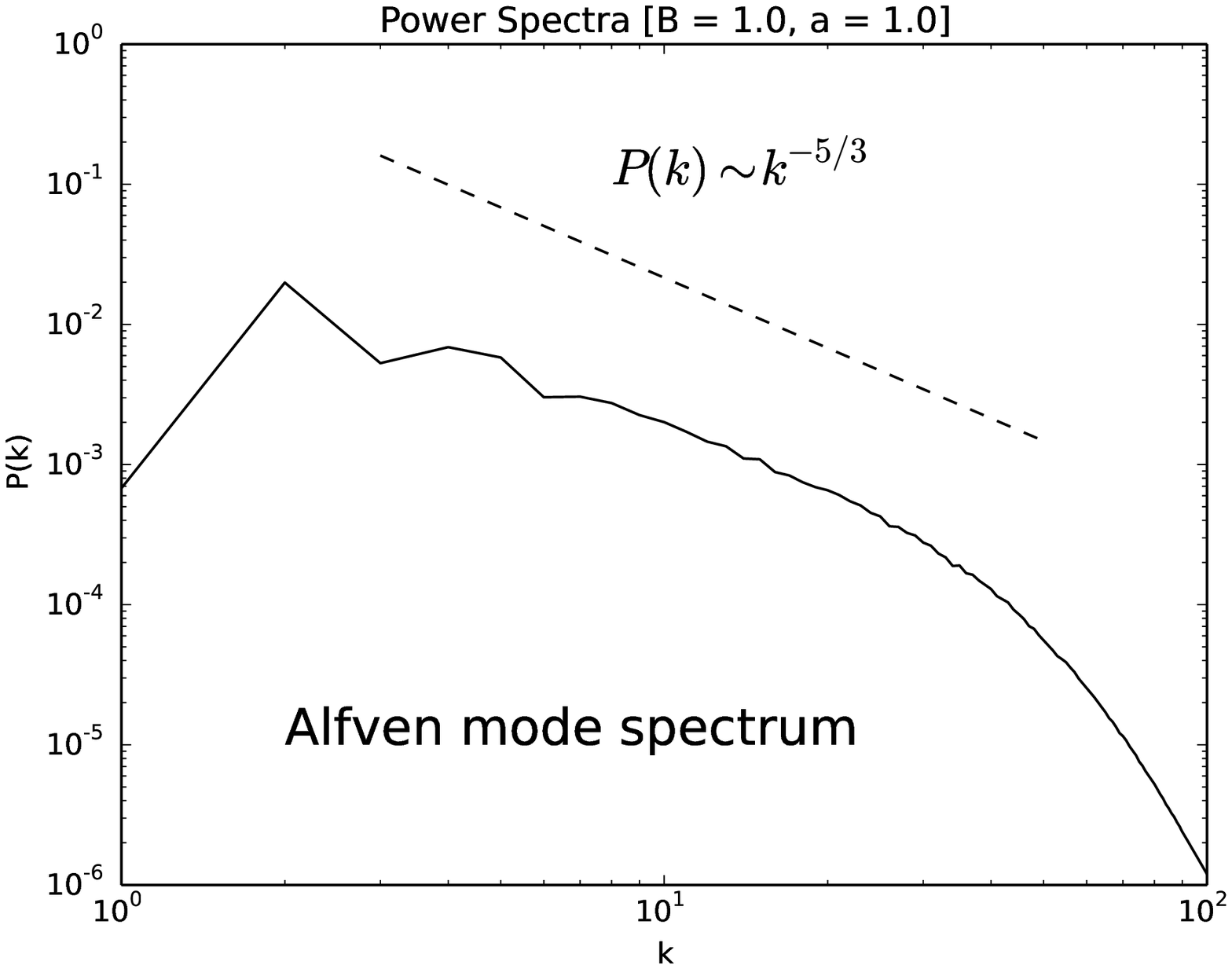}
\includegraphics[width=5cm]{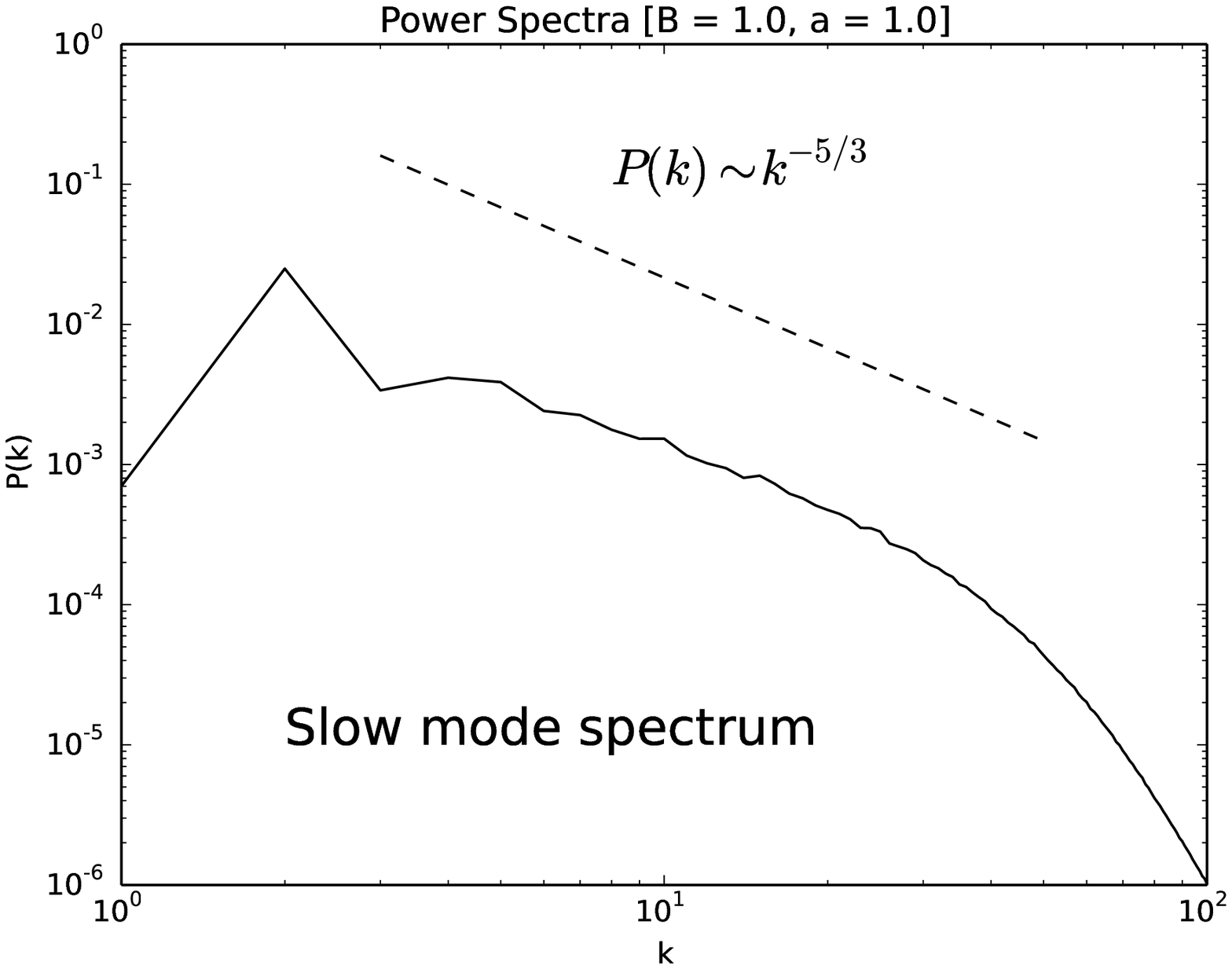}
\includegraphics[width=5cm]{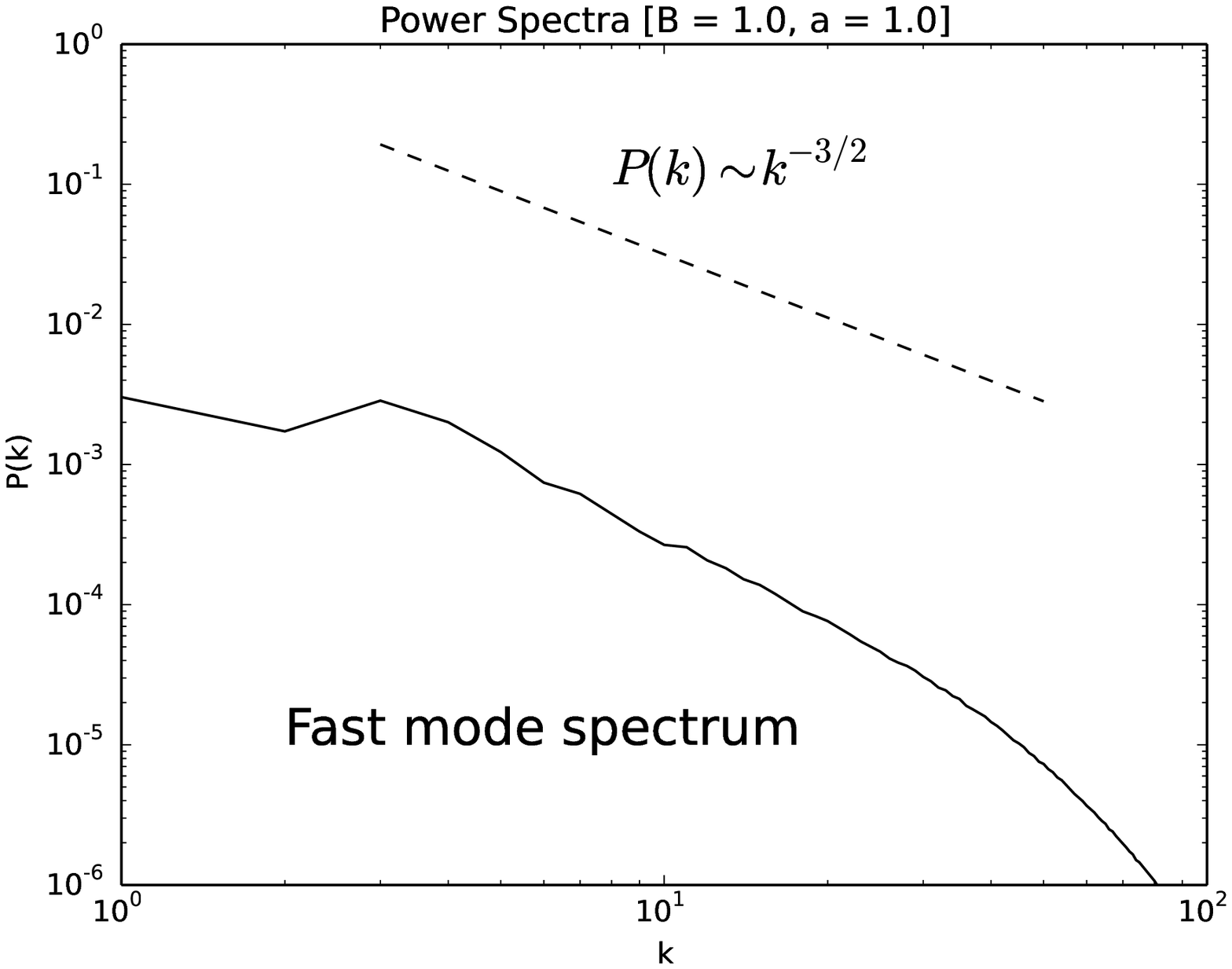}
\includegraphics[width=5cm]{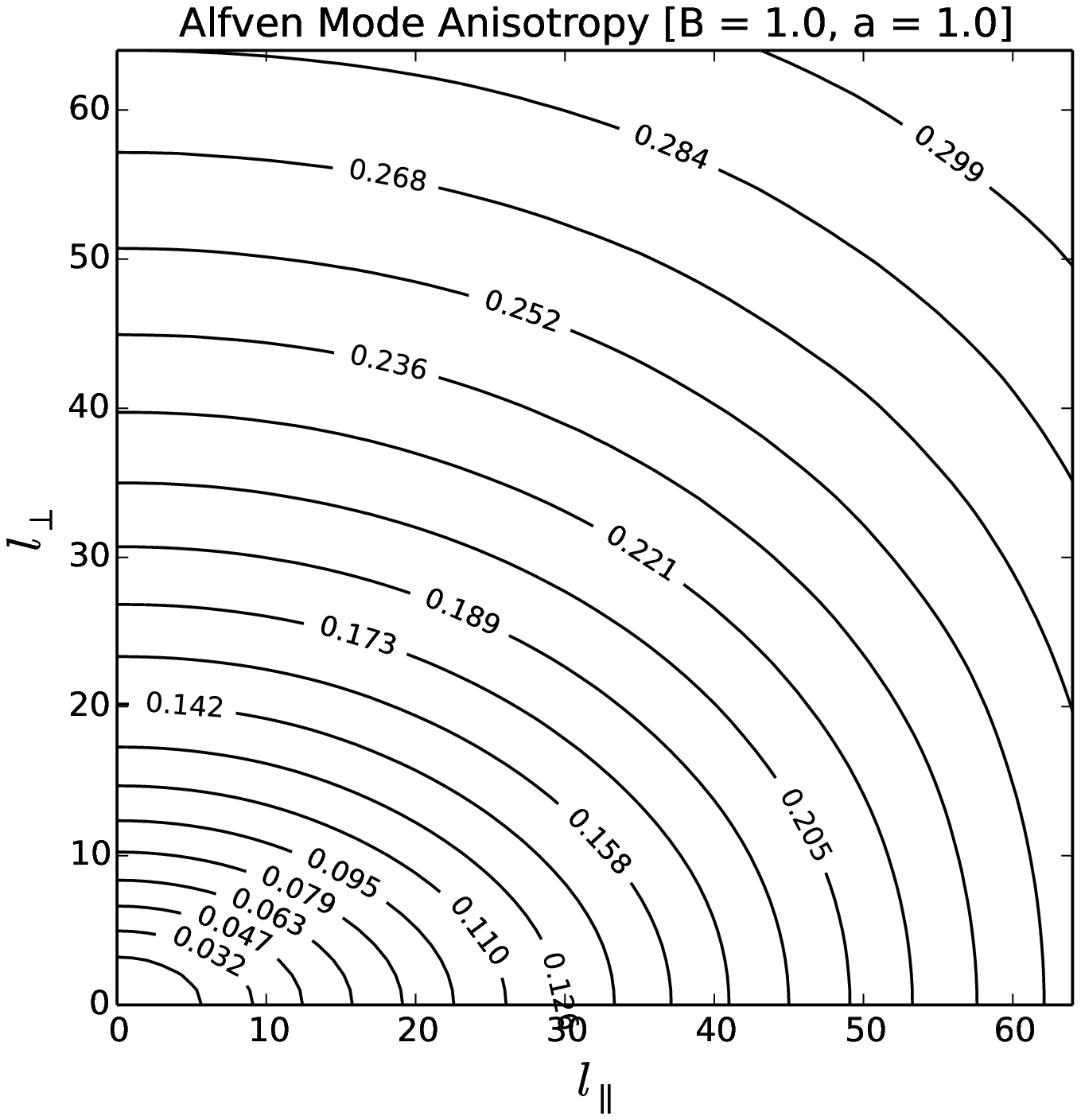}
\includegraphics[width=5cm]{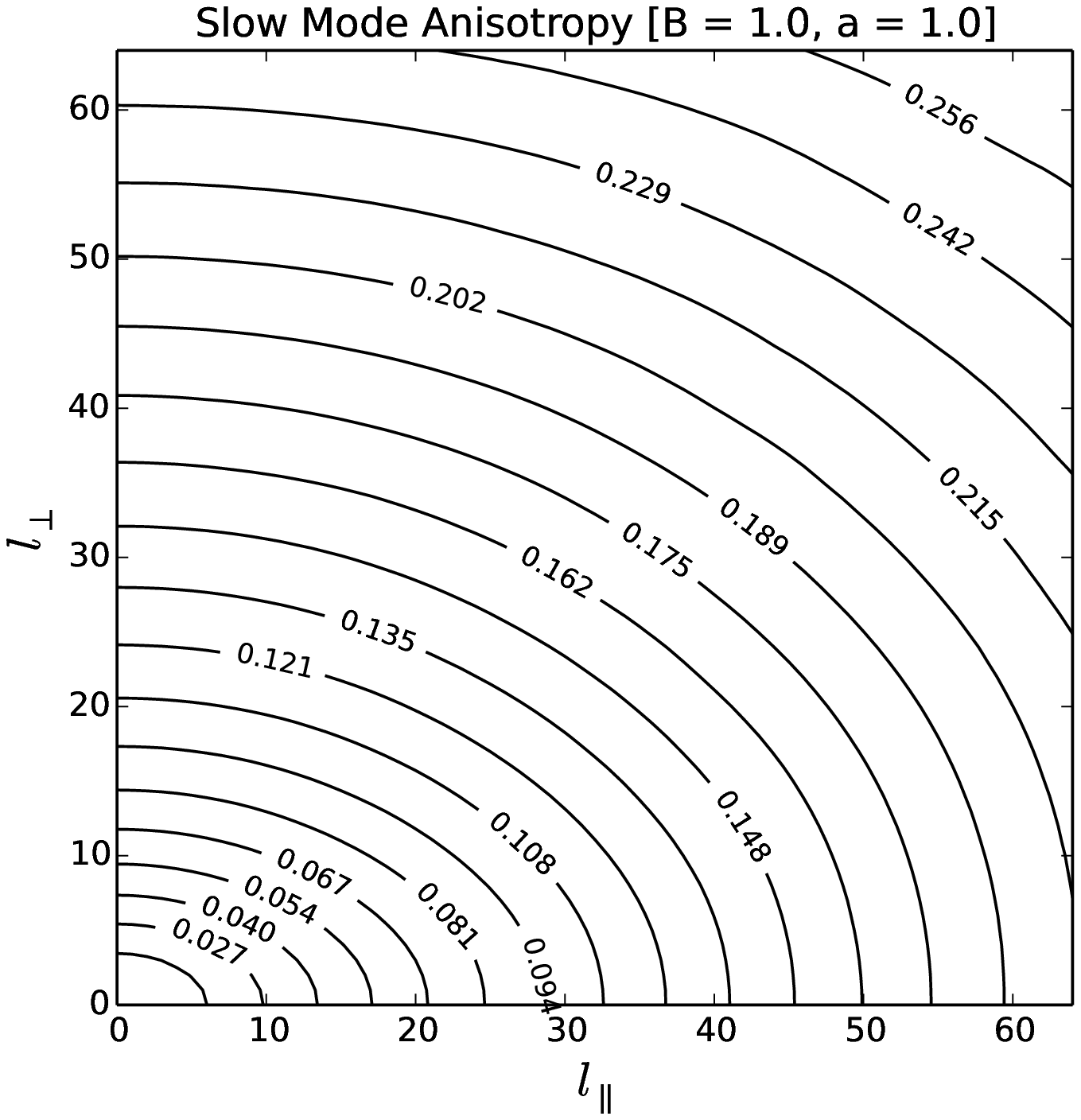}
\includegraphics[width=5cm]{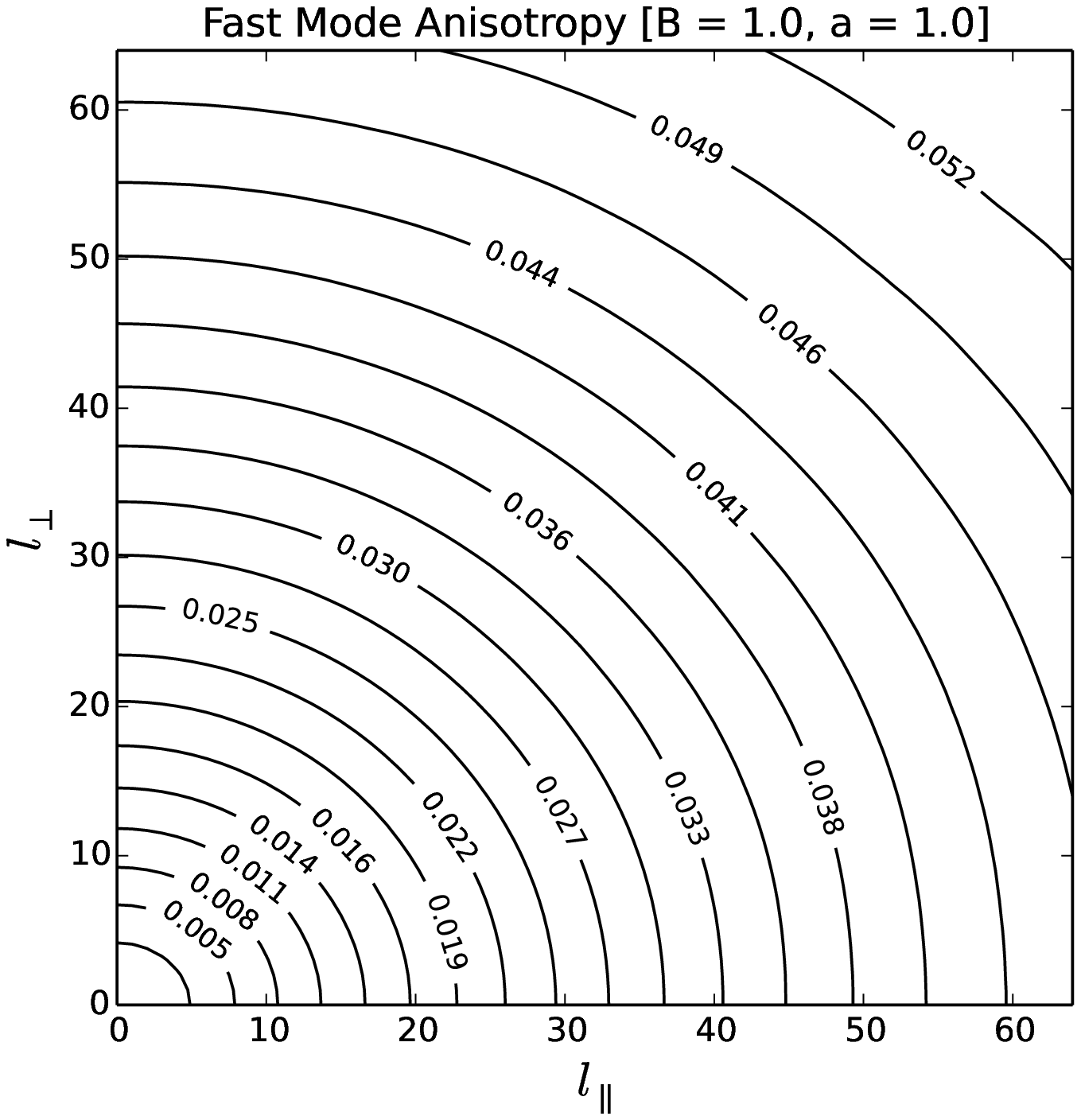}
\end{center}
\caption{Spectra and second order structure function anisotropy of 
dispersion ($\delta v$) of the different
wave modes in MHD turbulence.  The Alfv\'en and slow modes present K41 power
cascade and strong anisotropy of dispersion of velocity at small scales, while fast waves present IK
cascade and are basically isotropic at all scales. Data from a $1024^3$ isothermal, sub-Alfv\'enic and 
subsonic turbulence model. \label{fig2}}
\end{figure*}

\citet{mar90} remarked that if, instead of an Alfv\'en time, the timescale for
the waves to non-linearly interact with each other was the regular eddy turnover
time, i.e.  $\tau_l \simeq \tau_{\rm eddy} \sim l_{||}/\delta u_l$, 
one would get a K41 cascade for the
magnetized turbulence.  This would be true also for the case of strong turbulence,
$|{\bf z}^\pm| > \breve{B}_0$. The isotropy condition was retained, which was raising a
problem, most of the observational data
mentioned above revealing strongly anisotropic turbulence.

\citet[GS95 hereafter]{gol95} proposed a turbulent model based on anisotropic
fluctuations, with strong coupling between the wave modes.  Strictly speaking
the GS95 model assumes a critical balance between mechanical and
Alfv\'enic modes in such a way that $l_{\perp}/\delta u_l \simeq
l_{||}/v_A$.  
Therefore:

\begin{equation}
l_{||} \sim v_A \epsilon^{-1/3} l_{\perp}^{2/3}\\ {\rm and}\\
P_u(k) \propto k^{-5/3}.
\label{eq18}
\end{equation}

From Eq.~\ref{eq18}, not only the magnetized turbulence is anisotropic but it is
{\it local} in the sense that the anisotropy is measured in the reference frame
of the local magnetic field.  Such an anisotropy is expected to occur in both 
the dispersion of velocity ($\delta v$) and wave vectors ${\bf k}$, though it 
is easier to observe velocity dispersion anisotropies from the interstellar medium, as 
discussed below. 
Therefore, statistically, a large number of eddies
with local fields randomly distributed in space result in an average zero
anisotropy (even at small scales).  In the strong magnetized cases though, the
anisotropy would be more clearly detected in experiments and observations.

Several direct numerical simulations of magnetized turbulence in a
quasi-incompressible regime have been performed in the past decade.  Many
numerical experiments reveal that MHD turbulence indeed has a large part of its
energy cascade close to a K41 distribution.  However, as shown by
\citet{cho02a, cho03, cho02b} and \citet{kow10}, the decomposition of the
different modes in MHD turbulence actually reveals that, although Alfv\'en and slow
modes behave as K41 type of turbulence and are anisotropic, the fast modes are
isotropic and follow IK statistics (see Figure~\ref{fig2}).

Effects of imbalanced (or cross-helicity) turbulence in the cascade and
statistics of the local fields have also been addressed in the past few years
\citep[][and references therein]{lit07, ber08, ber10, wic11, mar13}.  Imbalanced turbulence occurs when waves
traveling in opposite directions along the mean magnetic field are of unequal
amplitudes, i.e. carry different energy fluxes to small length scales, so that
${\bf z}_l^+/{\bf z}_l^- \neq 1$ and  ${\epsilon}_l^+/{\epsilon}_l^- \neq 1$.
The imbalance may arise in MHD turbulence since the interaction timescales
between the waves ${\bf z}_l^+$ and ${\bf z}_l^-$ are different, and the cascade
generally occurs faster for ${\bf z}_l^-$. This is understood as the number 
of interactions ($N$) is much larger for counter-propagating wave packets, resulting in
${\epsilon}_l^+/{\epsilon}_l^- > 1$. 
 In such a scenario, numerical simulations
show that the anisotropy is not equal for the different wave modes.

Locality of scales for wave-wave interactions has also been the subject of recent
studies in turbulence \citep{car06, ale07, min08, alu09, ber10}.  Magnetic
fields are responsible for long range interactions, from the Lorentz force acting
over the whole fluid frozen to it.  Therefore, different wavelengths may interact with each
other non-linearly.  Bi-spectra of fluctuations of density are discussed in
\citet{bur09}, and the non-local interactions appear to be important in MHD
and supersonic turbulence models.  A similar approach is used for studying the
non-local interactions of Els\"asser modes \cite{cho10}, resulting in a
substantial fraction of non-local interactions in MHD turbulence.  The role of
the non-local interactions in the turbulent cascade is still not clear though.
Turbulence in magnetized collisionless plasmas has been also studied
in the past few years \cite[e.g.][and others]{hel06, sch08, bal09} in order to 
determine the role of collisionless plasma instabilities on the dynamics of 
plasma turbulence. Simulations of \citet{kow11, san13}, reveal that the statistics 
are still dominantly Kolmogorov-like, though strong asymmetries may also arise due to
instabilities (firehose, mirror and cyclotron instabilities) are small scales.

\section{Signatures of a turbulent ISM}

In the previous section some theoretical aspects of 
turbulence have been presented. Its direct comparison to the 
dynamics of the interstellar gas is not trivial, as we discuss in 
the following. However, we will present here some observational 
evidences for a turbulent ISM, and discuss the possible turbulent 
regimes that may be inferred from these.

The recognition of a turbulent interstellar medium dates back to
1950's with the work of \citet{von51} on the
spatial distribution of dense structures in the plane of the sky.  He
recognized the hierarchy of structures and suggested its turbulent 
origin.  The identification of turbulent motions
was provided shortly after it was measured from velocity dispersions
\citep{vonH51}.  Later on, the observational and 
theoretical supports for a turbulence dominated ISM have grown considerably
\cite[see reviews by][and references therein]{elm04, mac04, hen12}, causing 
a major shift in the uderstanding of the ISM nature, from a
thermal pressure dominated system, as thought before, to a very dynamic
multi-phase system.

\subsection{Density distributions}

As mentioned above, one of the main signature of the turbulent
character of the ISM is related to the density distribution of its contents. Up to now, 
tracers of the gas density distributions of the ISM at large scales have been dominantly indirect\footnote{{\it in situ} data have been obtained at the nearby interstellar plasma 
by Voyager 1 \citep{gurnett13}, though no direct study of the local turbulence has been discussed yet.}. They rely on spectral lines and continuum emissions from the different phases  
of the ISM: the hot and fully ionized (HIM), 
the warm and fully/partially ionized medium (WIM/WNM), and the
cold weakly ionized (CNM). These emissions being integrated along lines of sight and projected 
in the plane of the sky, sophisticated inversion methods have to be implemented.
It is the statistical approach of the 
temporal and spatial variability of these emission fluxes that are 
the readily accessible observational techniques for studying interstellar turbulence.

\begin{figure}[ht]
\vspace*{2mm}
\begin{center}
\includegraphics[width=8cm]{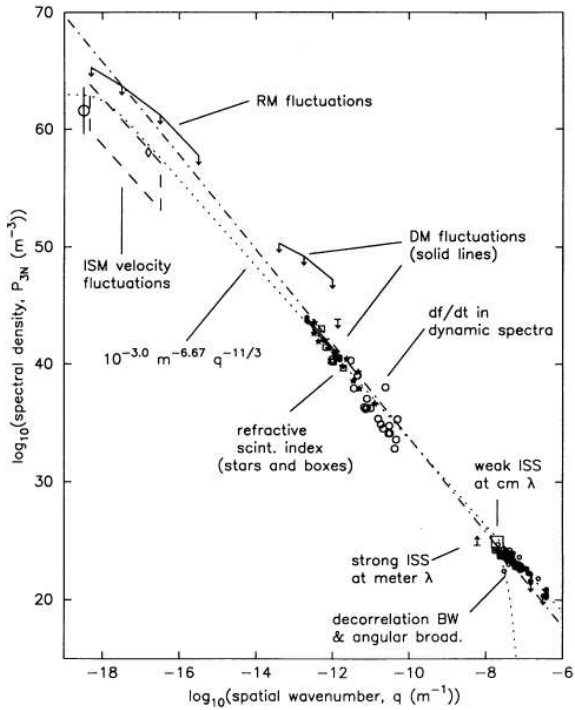}
\end{center}
\caption{Power spectrum of density along the line-of-sight from different data
sets, and the dashed line as reference for Kolmogorov-like spectrum for one
dimension ($k^{-11/3}$). \cite[extracted from][]{arm95}. \label{fig3}}
\end{figure}

With hydrogen being the most abundant element in the Universe, the $\lambda$ 21cm line of neutral 
hydrogen is a key diagnostic. Its line integrated emission is proportional to the bulk of the 
hydrogen column density, since its opacity remains low over most of the ISM.
Statistics of the HI intensity spatial distributions have therefore been used
to probe interstellar turbulence but the results are far from homogeneous. 
 \citet{gre93} studied the
power-law of the spatial power spectrum of the HI emission from different
fields in our Galaxy.  He obtained power spectra with slopes between -2.2 and
-2.8 at a scale range between 35 and 200~pc. From the HI 21 cm absorption
towards Cas A. \citet{roy09} derived a power law with index -2.7, 
consistent with Kolmogorov turbulence
in the diffuse interstellar medium. 
However, \cite{miv03} find an impressive power-law in the nearby ISM at high galactic latitude 
with the same slope of -3.6 over two orders of magnitude in scales (between 0.1 and 25~pc). 
Similar studies have been performed since then, including other density tracers 
such as the CO and $^{13}$CO line 
emission of molecular clouds and power-laws have also been inferred \cite[e.g.][]{ben01, hil08}. 
A review of the scatter of the power-law slopes measured is given in \cite{hen12}.
The scatter of the slope values is certainly affected by projection effects: 
one would expect a 2D power spectrum $k^{-8/3}$ for an intrinsic Kolmogorov scaling. However,
the integration along lines of sight crossing often large amounts of turbulent ISM 
with different properties tends to blur
such a simple law. Moreover, the different tracers originate  in truly different phases 
of the ISM with varying amounts of small scale structure that may affect the power spectrum 
of the density distributions (i.e. in many cases, like supersonic turbulence,
density fluctuations are not simply advected by turbulence as  passive scalars, 
see \citet[e.g.][]{aud05}).
Indeed, as seen in Fig. 10 of \cite{hen12}, many studies 
(including the power spectrum of the dust thermal emission) give power-laws indices  
close to -2.7. It is not possible though to presume that a Kolmogorov-like cascade 
operates in the ISM, with scalings given by Equations 2 and 3. Even though compressibility 
seems to play little effect on the statistics of the ISM, except for small scales ($\sim$ pc scales) and 
cold and dense regions, magnetization effects may be important, as we discuss further below.

 \citet{arm95} used another tracer of density fluctuations, the scintillation of the 
background radiation (i.e. changes in the refraction index due to the turbulent motions in the
ionized components of the ISM)  in order to obtain the density spectrum along the line-of-sight.
 As a complementary method, fluctuations of the 
Faraday rotation  measurements (RM)  in the plane of sky are
also used to estimate density fluctuations (once the magnetic field is known)
on the line-of-sight \cite[e.g.][]{min96}.  The combined data provide the
density fluctuations along the line-of-sight, but for different lengthscales, as
seen in Figure~\ref{fig3}.  The turbulence probed by both methods
(scintillations and RM) present a most impressive spectrum, with a unique 
Kolmogorov-like slope across more than ten orders of magnitude in wavenumber.

Similar works have been done for external galaxies.  Turbulence has 
been characterized based on similar techniques for the Small
Magellanic Cloud \cite[see][]{sta99, sta01, che08, bur10} and revealed
spatial variations of HI morphology.  \citet{dut13} calculated HI intensity
fluctuation power spectrum for a sample of 18 spiral galaxies and found slopes
in the range of -1.9 to -1.5.  Shallower spectra, compared to K41, could be evidence for
two-dimensional eddy dominated turbulence at scales larger than the disk thicknesses.

\subsection{Velocity fields}

\subsubsection{Direct statistical analysis}
\begin{figure*}[ht]
\vspace*{2mm}
\begin{center}
\includegraphics[width=16cm]{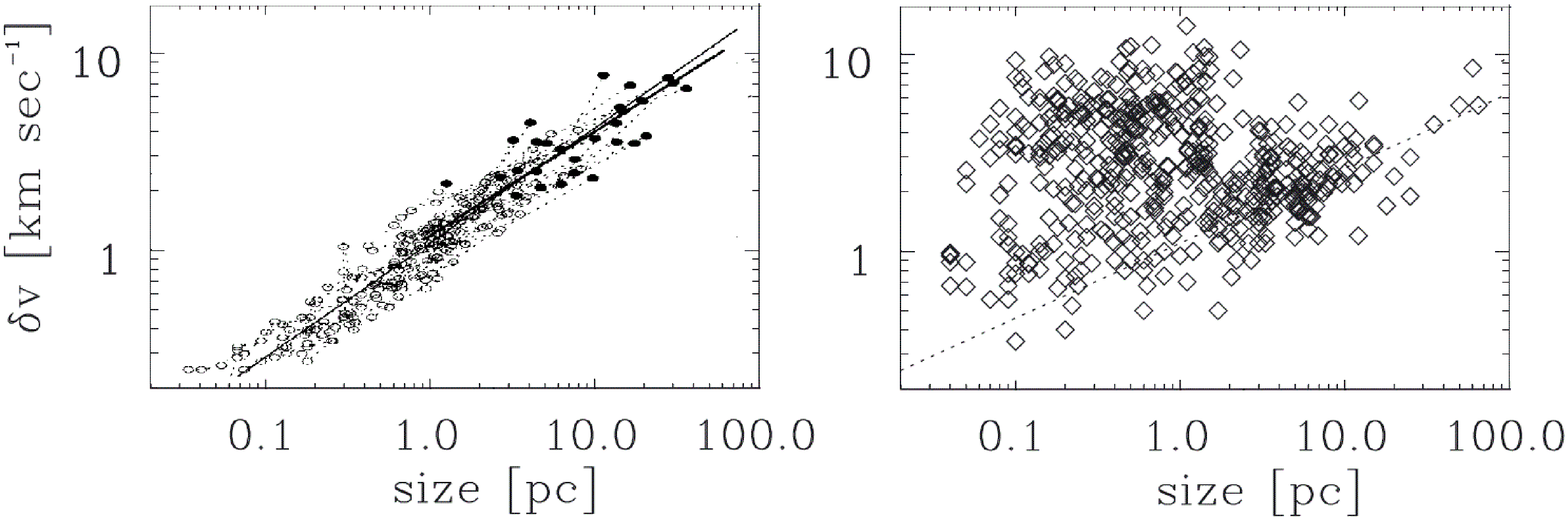}
\end{center}
\caption{Velocity dispersion relations from different surveys, \citet{hey04}
(Left panel) and compilation from several surveys done by \citet{bal11}
(Right panel). As the later authors point out, while large CO clouds from 
the survey by Heyer et al. (2009) exhibit the typical Larson relationship, denser structures 
show larger dispersion of velocity. This fact has been interpreted by those authors as 
due to gravity in collapsing cores, while \citet{fal10b} argued for projection effects and compressibility. \label{fig4}}
\end{figure*}

Spectral lines of several species observed with high spectral resolution  may be used to
infer the turbulence velocity distributions in the different phases of the ISM, such
as hydrogen lines (mostly) and some ions for the diffuse ISM
\cite[e.g.][]{bow08}, and molecular spectral lines ($^{12}$CO and $^{13}$CO in
most surveys) for the molecular clouds.  The early surveys of \citet{lar81} and
\citet{sol87} revealed the  universal line-width and mass distribution scalings 
among  molecular clouds.  Notably, both works
pointed to a velocity dispersion relation $\sigma_v \propto l^\alpha_\nu$, with
$\alpha \sim 0.5$ (see Figure~\ref{fig4}, left).  Many similar studies
were carried out to study the velocity distribution in molecular clouds, such as
the work by \citet{gol08, yos10, qia12, hey12} in the Taurus Molecular Cloud;
\citet{gus06} and \citet{liu12} for the Orion Complex, and many others.

More recent studies confirmed the same scaling relation
although with slopes varying
significantly \citep{hey04, qia12}.  \citet{qia12} for instance used
the variance of the velocity difference of cores in molecular clouds, instead of
the line width, and obtain $\alpha_\nu \sim 0.7$.  On the other hand, massive cores
are known to exhibit shallower slopes compared to what is frequently assumed
(i.e. $\alpha_\nu < 0.5$).

Recently, \citet{bal11} compiled different observational surveys and concluded
that while in general terms, the typical CO clouds observed by
\citet{hey09} lie close to Larson's relation, this is clearly not the case for
the dense and massive cores, which exhibit large velocity dispersions for their
relatively small sizes (Figure 4).  Those authors propose that the large dispersion
observed at small scales are related to increased velocities as the clouds
become gravitationally bound.  However the increased dispersion at small scales has
already been reported though, based on numerical simulations, in \citet{fal10}
without self-gravitating objects.  For these authors the large dispersion
observed at small scales is an intrinsic feature at the turbulent gas.  
The broad dispersion 
of the scaling relation indicates a turbulent regime dominated by compressible 
motion at small scales, as discussed in Section 1.1, though regular incompressible
turbulence dominates at larger scales. Compressibility, as 
described in Fleck's model (see Equation 10), naturally give larger slopes for 
the dispersion relation, with a value of $\alpha \sim 0.16$ favoured by observations. 
It is
not clear though what is the actual role of gravity in the statistics of the molecular
cloud emissions.

At the large scale end of the cascade, the apparent uniqueness of the scaling of the velocity dispersion 
with size scale suggests an universal source (or mixture of sources) 
of energy for the molecular gas turbulence in our
Galaxy.  \cite{che10} presented statistical analysis of high-latitude HI
turbulence in the Milky Way based on velocity coordinate spectrum (VCS)
technique.  They found a velocity power spectrum $P_u(k) \propto k^{-3.8}$
and an injection scale of $\sim 140 \pm 80$pc.  The alightly steeper slope, compared to
K41, can be the result of shock-dominated (compressible)
turbulence, with averaged sonic Mach numbers $\sim 7 - 8$ (see 
Section~\ref{sec1.1} above).

Two-point statistics are also used but, since in situ measurements are not yet available, one  
easily accessible observable turns out to be the variations in the plane-of-the-sky of the 
line-of-sight centroid velocity of spectral lines. \citet{lis96} 
showed that they trace the plane-of-the-sky projection of the vorticity.    
Using a sample of about one million independent CO spectra in a diffuse field, 
\citet{hil09} identified, on statistical grounds, the ensemble of positions 
at which vorticity departs from a Gaussian distribution. These form coherent elongated 
structures at the parsec-scale that are found to harbor sub-structures of  
most intense velocity shears down to the milliparsec scale \cite{fal09}. 
These coherent structures are proposed  to be the manifestations of the intermittency 
of turbulent dissipation in diffuse molecular clouds (see the review of \cite{hen12}),
which may be compared to Equation 8 above.

\citet{li08} studied the scaling relations of the velocity dispersions from
different neutral and ionized molecular species, namely HCN and HCO$^+$, in the
region of M17.  As it occurs in many other star forming regions, the ionized
molecules systematically present smaller dispersion of velocity compared to the
neutral.  Such a difference arises as turbulent energies dissipate differently
for the species due to ambipolar diffusion.  \citet{fal10b} showed that the
dispersion for ions is typically smaller than that for the neutral
species basically due to the damping of the ion turbulence at the
ambipolar diffusion scales ($\simeq 0.01$pc).

The direct comparison between statistics of observational data and the theory
must be done with caution.  Column density projections, or in
other sense emission maps, are influenced by projection effects.  Different
structures projected on  the same line-of-sight, but decorrelated at a given
lengthscale, may be observed as a single structure in the projected emission map.
Some deconvolution is possible though once the velocity profile is known, with high
spectral resolution.

\subsubsection{Indirect access to the velocity field via maser emission}

The low surface brightness of the above tracers 
and projection effects make the direct analysis of
turbulent flows in the ISM difficult.  Maser spots that are bright point sources and 
are transported by turbulence as passive scalars (because they are tiny and low mass structures), 
turn out to be powerful tracers of  the turbulent velocity field. 
Maser radiation in molecular lines appear in dense regions
where population levels inversion can be generated by radiative pumping, for instance,
 e.g. in the dense
molecular gas of star-forming regions (SFRs) associated with ultra-compact HII
regions, embedded IR sources, hot molecular cores, Herbig-Haro objects, and
outflows \citep{lit74,rei81,eli92,lo05}.  Maser emissions are often
characterized by high brightness temperatures and high-degrees of polarization.
Intense maser emission is detected in the molecular lines of hydroxyl
(OH), water (H$_2$O), silicon monoxide (SiO), ammonia (NH$_3$), methanol
(CH$_3$OH), among others.

\citet{wal84} used the {\it Very Long Baseline Interferometry} ($VLBI$) maps of
the H$_2$O maser source in W49N to demonstrate that both two-point velocity
increments and two-point spatial correlation functions exhibit power-law
dependencies on the maser spot separation, which is indicative of a turbulent
flow.  \citet{gwi94} performed statistical analysis of VLBI data for
W49N to confirm the power-law dependence of the velocity dispersion and spatial
density of masing spots on spatial scale, and interpreted this observation as an
evidence of turbulence.  \citet{ima02} reported sub-milliarcsec structures of
H$_2$O masers in W3 IRS 5.  
A cluster of maser spots (emission spots in
individual velocity channels) displays velocity gradients
and  complicated spatial structure.  Two-point spatial correlation functions
for the spots can be fitted by the same power laws in two very different spatial
ranges. The Doppler-velocity difference of the spots as a function of spot separation
increases as expected in Kolmogorov-type turbulence. \citet{str02} used VLBI data to
investigate the geometry and statistical properties of the velocity field traced
by H$_2$O masers in five star-forming regions.  In all sources the angular
distribution of the H$_2$O maser spots shows approximate self-similarity
over almost 4 orders of magnitude in scale. The lower order structure
functions for the line-of-sight component of the velocity field can be fitted by
power laws, with the exponents close to the Kolmogorov value.
Similar results were also obtained for other
regions \cite[e.g.][]{ric05, str07, usc10}.

\subsection{Turbulent magnetic fields}

The magnetic fields in our Galaxy is modelled as a superposition of 
different components: i) a large scale field, following a  spiral structure
in the plane of the galactic disk, and extending high above the plane into the Galactic halo, 
and ii) a complex component of
locally disturbed magnetic fields, which are related to molecular clouds and
star formation regions.  The spiral pattern in the disk aligns with 
the spiral arms \cite[e.g.][]{han06}.  This is
expected since the shear of gaseous motion around the center of the
galaxy stretches the field lines in this direction \cite[see review of mean
field dynamo by][]{bec96}.

There are four main methods to study the fluctuations in the ISM magnetic field,
namely the polarization of dust thermal emission  (both in emission in the far-infrared 
(FIR) and absorption in the visible and near-IR), Zeeman
effect of spectral lines, Faraday rotation and  polarization of the synchrotron  
emission.  Polarized synchrotron 
emission can also be mapped in order to
provide the geometry of the field lines in the plane of the sky. Faraday
rotation and synchrotron polarization measurements excel in probing the magnetic
field of the diffuse ionized medium of the ISM, i.e. they are excellent tools to
study the large scale fields of galaxies in general. More extensive reviews both
on magnetic fields in star formation regions and galactic
scale magnetic fields are given in \citet{cru12} and
\citet{han06}, respectively.

As mentioned earlier, synchrotron emission polarization can
be used for mapping the large scale structure of the magnetic fields in galaxies
\cite[see review by][]{bec09}.  The fields traced by the polarized synchrotron
emission present intensities of the order of $\sim 10 - 15\mu$G.
However, the synchrotron emission probes the ionized medium
only, which is less useful in determining the
turbulence properties of the star formation regions, dominated
by the dense and neutral components of the ISM.  Therefore, a magnetic field with
intensity $\sim 10 - 15\mu$G is supposed to thread most of galactic disk,
except the dense regions of the arms where the local properties of
the plasma and stellar feedback may dramatically change the field properties.

\citet{opp12} compiled an extensive catalog of Faraday rotation measure
(RM) data of compact extragalactic polarized radio sources in order to
study the angular distribution of the all-sky RMs.  The authors found
an angular power spectrum $P(k) \propto k^{-2.17}$ for the Faraday
depth, which is  given by the product of the line-of-sight magnetic field 
component $B_{\rm LOS}$ and the electron number density $n_e$. The
combination of the RM and polarization vectors of the synchrotron emission
allows to reconstruct the three-dimensional structure of
galactic magnetic fields.  Such angular fluctuations of the Faraday depth is
thought to be related to the turbulent ISM.  However, the
relationship between the fluctuations of the RM and the local fluctuations of
electron density and magnetic fields is not clear yet.  This, for
instance, is an interesting subject for further comparisons
with simulations (as in \citet{gae11}).

Possibly the most direct method for estimating the magnetic field intensity in
the dense and cold ISM relies on the detection of Zeeman effect \cite[see][for
details]{rob08}.  For instance, \citet{sar02} detected and
studied the Zeeman effect in H$_{2}$O masers in several SFRs and
determined line-of-sight magnetic field strengths ranging from 13 to 49 mG.
They found a close equilibrium between the magnetic field energy and
turbulent kinetic energy in masing regions.

\citet{alv12} showed that shock-induced H$_{2}$O masers are important
magnetic-field tracers of very high-density gas in low-mass protostellar core
IRAS 16293-2422.  They investigated whether the collapsing regime of this source
is controlled by magnetic fields or other factors such as turbulence, and
concluded that the magnetic field pressure derived from data is comparable to
the ram pressure of the outflow dynamics.  This indicates that the magnetic
field is energetically important for the dynamical evolution of the
protostellar core.

Due to its brightness, maser emission is better for probing 
magnetic fields, but they are rare and limited in extent. 
The Zeeman effect in non-masing regions has been detected for
HI, OH, and CN lines for which the turbulent
broadening is typically larger than the Zeeman splitting in frequency. 
The compilation by \citet{cru99} and recent CN 
Zeeman observations in SFRs by \citet{falgarone08} show that the
turbulent motions within the SFRs and molecular clouds are supersonic but
sub-Alfv\'enic.  The {\it upper limit} magnetic field intensity  scales
with density, estimated from a Bayesian analysis, as 
 $B \propto n^\kappa$, with $\kappa \sim 0.47$ \citep{cru10}.  Collapsed
structures along the mean field would produce $\kappa \rightarrow 0$, while
shock compressions perpendicular to the field lines result in $\kappa
\rightarrow 1$.  The observed relation with $\kappa \sim 0.47$ is expected, for instance,
in Alfv\'enic perturbations  and is in agreement
with MHD simulations \cite[e.g.][]{bur09}.  It was also claimed in
that work that, despite its relative importance in the overall dynamics of
clouds, the uniform magnetic fields in these clouds are in general not strong
enough to prevent gravitational collapse based on the mass-to-flux ($M/\Phi$)
ratios observed. Other major compilations of Zeeman measurements in molecular
clouds are given, for example, in \citet{bou01} and \citet{tro08}
with similar results.

\subsubsection{Polarization maps of molecular clouds}

Radiation may be polarized due to a prefered direction for emission/absorption
from aspherical dust grains, as well as by some molecules and atoms.  The ISM in
known to be populated by a complex distribution of grain sizes and shapes.
Depending on its composition an aspherical rotating dust particle may align
with the magnetic field line.  The orientation of the polarization of
radiation is then linked to the orientation of the magnetic field itself
\cite[see review by][]{laz07}. Many observational data has been made available 
in the past decade both on absorption and emission dust polarization 
\cite[e.g.][]{heiles00,chapman11}.

The strengths of magnetic fields can be
estimated from polarization maps by the \citet{cha53} (CF) technique.  The CF
method is based on the assumption that the magnetic and turbulent kinetic
pressures are the dominant ones within the cloud, and that the fluid motions are
coupled to the magnetic field lines.  In this sense, any perturbation from the
fluid turbulence will result in a change in the orientation of the field lines.
Major improvements on the CF technique are given e.g. by \citet{fal08},
\citet{hild09} and \citet{hou09}.  If the velocity dispersion $\delta v_{\rm
los}$ is known, e.g. from spectral lines, the mean magnetic field in the plane
of sky can be estimated as \citep{fal08}:

\begin{equation}
B_{\rm sky}^{\rm uniform} \simeq \delta v_{\rm los} \left(4 \pi <\rho> \right)^{1/2} \left[1+\tan^{-1} \left(\delta \phi \right) \right],
\end{equation}

\noindent
where $\delta \phi$ represents the dispersion in the polarization angle.

From the equation above, the ratio $\delta B/B_{\rm sky}$ - assumed to be 
$\sim \tan^{-1} \delta \phi$ - is directly related to
the Alfv\'enic Mach number of the turbulence.  Notice that the
dependence of the projected $\delta B/B_{\rm sky}$ with the actual 3D MHD
turbulence may be removed from higher order statistical analysis, as proposed in
\citet{fal08}.

The left image of Figure 5 presents the polarization map of the Muska Dark Cloud \citep{per04} 
in the optical wavelengths, as a result of dust absorption. Vectors represent the magnetic field 
orientation. The filamentary morphology of the dark cloud is perpendicular to the external field, 
which is very uniform indicating a sub-Alfv\'enic turbulent regime. At the right hand side of Figure 5, 
the polarization map overplotted on the column density projection of a 3D MHD numerical simulation of 
sub-Alfv\'enic turbulence \citep{fal08}. Such comparisons between MHD numerical simulations and measurements 
of magnetic fields in the ISM are important in unveiling the physics of MHD turbulence, and its role on other 
phenomena such as star formation.

Spatial dispersion of magnetic fields in molecular clouds from polarization maps
may be used to characterize the
power spectrum of magnetized turbulence in the inertial and dissipation ranges.
 \citet{hou11} found a power law inertial range for the magnetic field spatial
distribution that is $\propto k^{-2.9 \pm 0.9}$, and a cutoff at scales $\sim
0.009$pc, which is claimed by the authors to be related to the ambipolar
diffusion scales.

\begin{figure*}[ht]
\vspace*{2mm}
\begin{center}
\includegraphics[width=7cm]{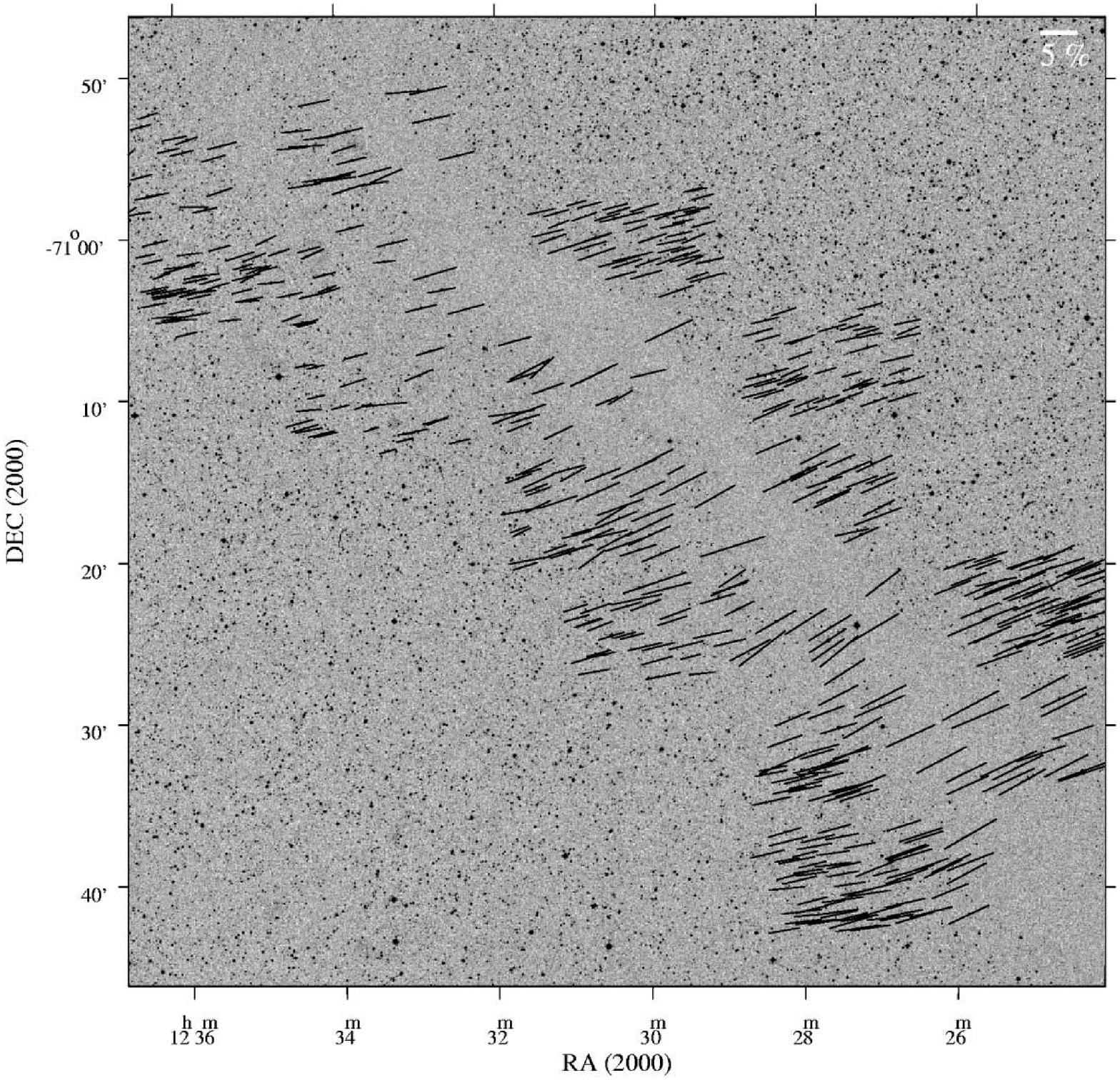}
\includegraphics[width=8.5cm]{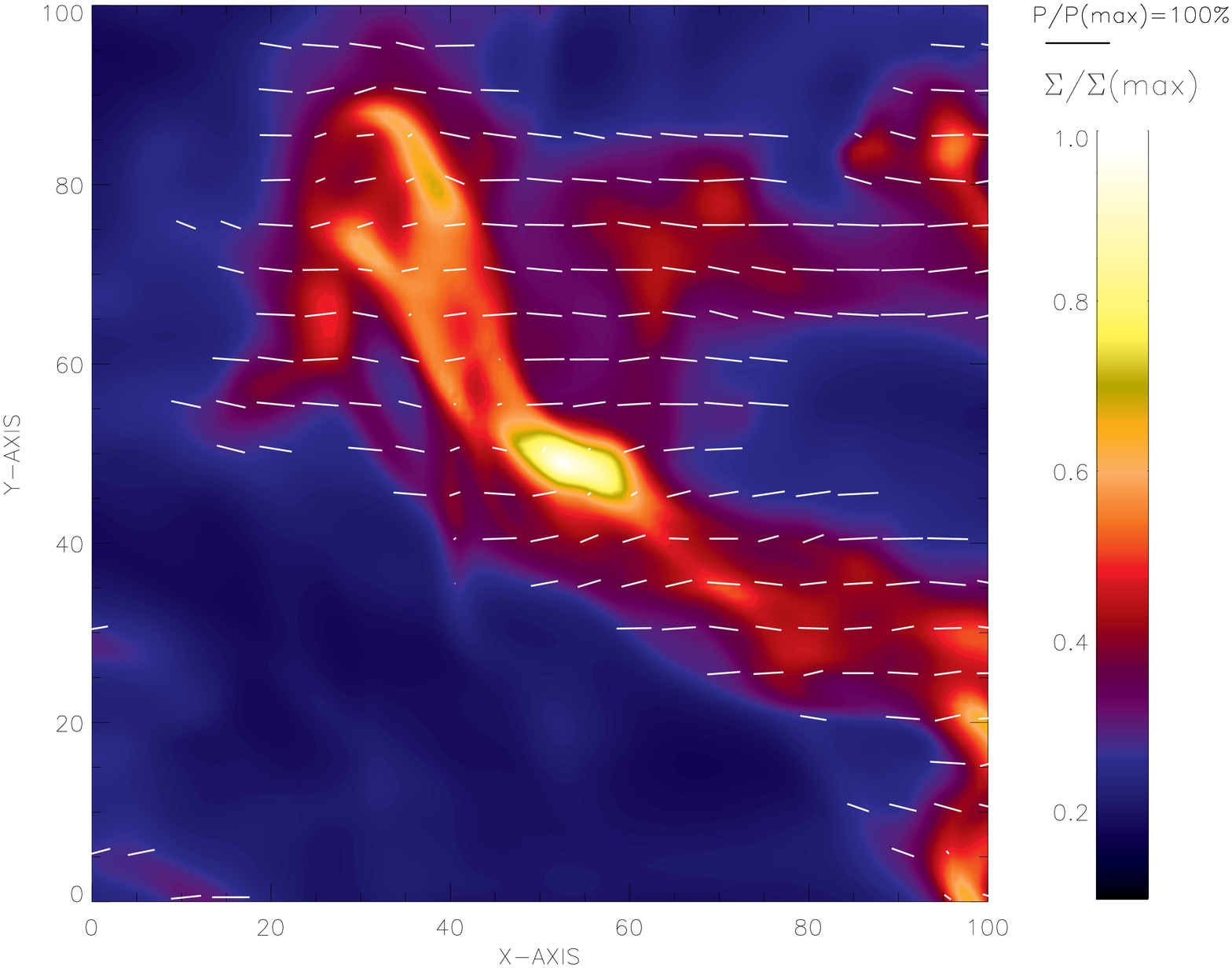}
\end{center}
\caption{Left: optical polarization map of the Muska Dark Cloud \cite[extracted from][]{per04}.  Right: simulated polarization map from a three dimensional simulation of MHD sub-Alfv\'enic turbulence \cite[extracted from][]{fal08}. \label{fig5}}
\end{figure*}

Again, as another issue in a proper modelling of the
statistics of velocity, gravity is claimed to interfere in the statistics of the
observed polarization maps \citep{koc12a, koc12b}.  Gradients in emission
towards the cores of molecular clouds have been shown to be associated to
gradients in the polarization angles.  A transition from a magnetically
subcritical to a supercritical state\footnote{i.e. the system becomes 
supercritical once gravity overcomes the magnetic pressure.}  
  could then explain the trend, and this
technique could provide an independent way to estimate the local
magnetic force compared to gravity.

\citet{hey12} showed that the turbulence in the densest regions of Taurus
molecular cloud is super-Alfv\'enic, while the reverse is true in the surrounding lower density
medium, threaded by a strong magnetic field.  This observational
result is in agreement with the transition expected between
scales as dense structures are formed, e.g. by shocks, in a supersonic but
sub-Alfv\'enic large scale turbulence \cite[see discussion in][]{fal08, hey08, bur09}.

Similar to the synchrotron radiation case, by combining
dust polarization maps with Zeeman measurements in molecular clouds one can
determine the three-dimensional structure of the magnetic field.  \citet{poi13}
recently succeeded in testing this approach for a number of objects of the
SCUBA Polarimeter Legacy (SCUPOL) data catalog.  The authors were able to determine the orientation of the
mean field with respect to the line-of-sight, as well as to estimate the
turbulence regime within several molecular clouds.  The authors also claimed
that all observed clouds seem to present a universal large scale turbulence that
is supersonic ($M_s \sim 6 - 8$) and sub-Alfv\'enic ($M_a
\sim 0.5 - 0.9$), at scales as large as 50pc.

In terms of comparing these data with basic theories of magnetized turbulence, 
most observations point towards a magnetically dominated turbulence at 
scales larger than few tenths of parsecs. \citet{hey08} also showed one of the 
first evidences for anisotropic turbulence in molecular clouds, with respect to the 
large scale magnetic field orientation. The observations of the Taurus Molecular Cloud 
revealed a significant anisotropy in the dispersion of velocity ($\delta v$), 
being larger for lags perpendicular to the mean large scale field lines. 
Even though a Goldreich-Sridhar similarity relation is not obtained, the anisotropy 
observed is a strong indication for strong coupling between MHD wave modes in the 
insterstellar turbulence, as predicted by GS95 model. 
We could extrapolate a bit and say that a GS95 model 
combined with fractal density distributions, as given in \citet{fle96}, 
is favoured for the ISM turbulence based on current observations.

\section{Origins of interstellar turbulence}

Surveys of different atomic and molecular line emissions have shown us
that the diffuse ISM is turbulent at scales $>150$pc, with $\delta v \geq 50$km
s$^{-1}$.  This results in a specific energy transfer rate\footnote{this
estimate is at least one order of magnitude larger than that of \citet{mac04},
since these authors considered a lower injection
velocity at the largest scales ($\delta v_L = 10$km s$^{-1}$).} of $\epsilon
\simeq m_H n_H \delta v_L^3/L \sim 10^{-25} - 10^{-24}$erg cm$^{-3}$ s$^{-1}$.
\citet{bru09} estimated the driving scales of turbulence for
molecular clouds by comparing observed and synthetic CO
velocity dispersions from numerical simulations.  They found that only models
with large scale sources of turbulence, such as supernovae-driven outflows (SNe) and galactic
dynamics, fit well to the observed data.

Supernovae have been claimed as main turbulence drivers by many authors
\cite[e.g.][]{nor96, mac04, avi04, avi05, jou06, hil12}.  It certainly
corresponds to an important driving mechanism for turbulence in starburst regions
and small galaxies \cite[e.g.][]{fal10, rui13}.  However, its impact on galactic
turbulence, in a more generalized sense, is still a matter of debate.

One issue is that numerical simulations of SNe driven turbulence  
create superbubbles that are far too hot and diffuse \cite[see][]{avi05, jou06, 
hil12}.  Other critical arguments disfavour SNe as a main
driver mechanism as well, at least for our Galaxy.  \citet{zha01} analysed the CO
emission lines from the Carina Complex and obtained a turbulent energy flux per 
mass density unit cascading over scales $\sim 10^{-7}$(km s$^{-1}$)$^2$ 
yr$^{-1}$, that could not
be explained from stellar feedback, but is in rough agreement
with the injection rate of energy from the gravitational interaction of
the ISM gas and the galactic spiral arms.  \citet{san07} also showed that HI mapping of
our Galaxy is consistent with a turbulence injection rate that is not directly
related to the star formation rate, but is about constant with respect to
the galactocentric radius.  Also, the correlation lengths related to SNe
turbulence is strongly dependent on local properties (such as local density and
temperature) \cite[see][]{lea09}.  Such local dependence also occurs with
respect to the height related to the galactic plane, since SNe
energy is easily released outwards \cite[e.g.][]{mel09, mara13}.  The 
universality of the observed properties of turbulence in our Galaxy, 
together with the extremely large
injection scales ($>100$pc) suggest a Galactic scale driving
source, which is later amplified, as second order effects, by local stellar feedback.  
\cite{qia12}, for instance,
obtained similar core and ambient turbulent statistics, which suggested that
molecular cores condense from more diffuse gas, and that there is
little (if not none) additional energy from star formation into the more
diffused gas.

Turbulence driven by galactic dynamics models, such as driven by
velocity-shears in the galactic disk, have been posed long ago \cite[e.g.][]{fle81}.
Instabilities such as the magneto-rotational instability have also been proposed
\cite[e.g.][]{sel99, kim02}.  Interactions between the arms of the Galaxy and
the disk gas also generate perturbations, as large as $20$km s$^{-1}$
\citep{gom02}, that could explain most of the injection of energy into turbulent
motions.
It is not clear yet which of these mechanisms (SNe or galactic
dynamics) is more important for the observed turbulence in the ISM.
Certainly, it is a promising subject for
studies in the upcoming years, both from theoretical and observational sides.


\conclusions  

In this work, we briefly reviewed part of the current understanding on 
incompressible, compressible and/or magnetized turbulence, which can be applied to 
characterize the interstellar medium. There is a
vast literature available for each of these and a complete review on turbulence is out 
of the scope of this work. We discussed the recent theoretical 
improvements made on the modelling and characterization of the different 
turbulent regimes. Multifractal description, statistics of probability functions, and spectral 
analysis are just a few that have been currently employed to characterize spatial 
and temporal variations of plasma properties associated to turbulent motions. 

Phenomenological descriptions of turbulence in Fourier space, such as that of 
Kolmogorov-Obukhov, are particularly simple and still very useful on the diagnostics 
of interstellar turbulence. Since scaling relations for compressible, incompressible and 
magnetized turbulence of these theories may differ among each other, observations 
can be used to determine the turbulent regime of the ISM. 

Spectroscopy has been long used to probe the velocity distributions along the line of sight. 
The observed amplitudes of the turbulent motions indicate that the ISM 
transits from a supersonic turbulent regime at scales of tens to hundreds of parsecs, 
at which the turbulence is driven, to subsonic at smaller scales. The scales where turbulence is 
subsonic depend on the ``phase"' of the ISM plasma. Dense molecular clouds present 
lower temperatures, which result in 
subsonic turbulence only at very small scales ($\ll 1$pc). The warmer 
and more diffuse media, such as warm neutral medium and the warm diffuse medium,
 present subsonic flows at scales of few parsecs due to the larger local sound speeds. 
 It is interesting to mention that this transition is deeply related to the 
 origin of the dense molecular 
clouds. These objects are either originated due to the large scale compressible motions 
of the gas \citep[e.g.][]{will00}, or at small scales due to other mechanisms, such as 
thermal instabilities. Current observations favour the first, given their lengthscales 
and internal dynamics \citep[see][]{poi13}.
Spatial gas distributions over the plane of sky are also provided observationally. The 
filamentary structure observed reveals a compressible dominated turbulent regime, at 
least at most of the observed scales. Observations also reveal a magnetized ISM. 
All these ingredients combined result in compressible and magnetized ISM turbulence, 
challenging theorists to provide a phenomenological description 
of the combined effects of supersonic flows and strong magnetic fields. Despite 
the good agreement between observations and the Goldreich-Sridhar model for magnetized 
turbulence, and Fleck's model for compressible one, such as spectral slopes, scaling 
relations and multifractal analysis applied to emission maps, a complete unified theory 
is yet to be developed. 

One of the major problems in comparing statistics of observed quantities to 
theories of turbulence relies on projection effects. 
Observations are spatially limited in the sense
that all statistics are done either along the light-of-sight (e.g.
scintillation, velocity dispersion from spectral lines, Faraday rotation)
or in the plane-of-the-sky. In addition, even the plane-of-sky maps are related to
integrated quantities (e.g. emission lines, column
density, Stokes parameters for the polarization maps). One
must therefore be careful when comparing these with theories 
of three dimensional turbulence.

Other effects may also make the direct comparison between theory and 
observations challenging.
Self-gravity of dense gas and stellar feedback, for instance, have been neglected in this paper. 
These processes are responsible for extra sources of energy and momemtum, but are not 
easily linked to the turbulent cascade. Despite their obvious importance on the 
process of e.g. star formation, their role on the statistics of the turbulence is not 
completely clear. Naturally, fragmentation and clumping would be enhanced if self-gravity 
is considered \citep{vaz96,bal11,cho11}, however its role on the cascade itself and on 
intermittency is unknown. 

Future studies from the theoretical side are possibly to be focused on the understanding 
of combined processes, such as magnetic fields, gravity, compressibility and radiation, on 
the energy transfer among scales. Formation of coherent structures and how their statistics 
relate to the bulk of the fluid are vital for theories of star formation. 
New data is also expected for the upcoming years. Although the
Herschel mission ended in early 2013 its data are not yet fully explored. 
Other major observational facilities, such as 
the Planck\footnote{Planck mission main goal is to observe the comic microwave background emission, for cosmological purposes, however the foreground ambient is the ISM, and proper modelling of the ISM structure and magnetic fields will be mandatory.} satellite and the Atacama Large Millimeter Array (ALMA), will provide complementary data at radio and microwave frequencies with very large sensitity, therefore going ``deeper" than 
reached by other instruments. Also, \citep{gurnett13} 
recently presented the first {\it in situ} measurement of the interstellar plasma as 
Voyager 1 has crossed the heliopause and started to probe the nearby interstellar plasma. This opens 
new possiblities in studying interstellar turbulence locally. It is clear that the future of 
interstellar turbulence science is going to be very exciting.

\begin{acknowledgements}
DFG thanks the European Research Council (ADG-2011 ECOGAL), and
Brazilian agencies CNPq (no. 300382/2008-1), CAPES (3400-13-1)
and FAPESP (no.2011/12909-8) for financial support. GK acknowledges support
from the FAPESP grant no. 2013/04073-2. ACLC thanks the European Commission for the award of an Marie Curie International Incoming Fellowshp, CNPq for support, and Paris Observatory for the kind hospitality.

\end{acknowledgements}







\end{document}